# Optimal Scheduling of Integrated Demand Response-Enabled Community Integrated Energy Systems in Uncertain Environments

Yang Li, *Senior Member, IEEE,* Bin Wang, Zhen Yang, Jiazheng Li, Guoqing Li

*Abstract*—The community integrated energy system (CIES) is an essential energy internet carrier that has recently been the focus of much attention. A scheduling model based on chance-constrained programming is proposed for integrated demand response (IDR)-enabled CIES in uncertain environments to minimize the system operating costs, where an IDR program is used to explore the potential interaction ability of electricity-gas-heat flexible loads and electric vehicles. Moreover, power to gas (P2G) and micro-gas turbine (MT), as links of multi-energy carriers, are adopted to strengthen the coupling of different energy subsystems. Sequence operation theory (SOT) and linearization methods are employed to transform the original model into a solvable mixed-integer linear programming model. Simulation results on a practical CIES in North China demonstrate an improvement in the CIES operational economy via the coordination of IDR and renewable uncertainties, with P2G and MT enhancing the system operational flexibility and user comprehensive satisfaction. The CIES operation is able to achieve a trade-off between economy and system reliability by setting a suitable confidence level for the spinning reserve constraints. Besides, the proposed solution method outperforms the Hybrid Intelligent Algorithm in terms of both optimization results and calculation efficiency.

NOMENCLATURE

**Acronyms**

| | |
|---|---|
| RG | Renewable generation |
| P2G | Power to gas |
| MT | Micro-gas turbine |
| PV | Photovoltaic |
| WT | Wind turbine |
| CIES | Community integrated energy system |
| HSD | Heat storage device |
| ESD | Electricity storage device |
| EB | Electric boiler |
| EV | Electric vehicle |
| IDR | Integrated demand response |
| CCP | Chance-constrained Programming |
| SOT | Sequence operation theory |
| MILP | Mixed-integer linear programming |
| PDF | Probability density function |
| TSE | Time-shiftable electric load |
| IE | Interruptible electric load |
| TSQ | Time-shiftable gas load |
| IQ | Interruptible gas load |
| CH | Cuttable heat load |

**Variables**

| | |
|---|---|
| $P^{WT}$ / $P^{PV}$ | WT/ PV output power (kW) |
| $P_{load,t}$ | Actual electric load in period $t$ (kW) |
| $Q_{load,t}$ | Actual gas load in period $t$ (m³) |
| $H_{load,t}$ | Actual heat load in period $t$ (kW) |
| $P_t^{TSE}$ / $P_t^{IE}$ | Time-shiftable/ interruptible electric load in period $t$ (kW) |
| $Q_t^{TSQ}$ / $Q_t^{IQ}$ | Time-shiftable/ interruptible gas load in period $t$ (m³) |
| $H_t^{CH}$ | Cuttable heat load in period $t$ (kW) |
| $P_t^{EB}$ | Absorbed electric power of EB in period $t$ (kW) |
| $H_t^{EB}$ | Heat output of EB in period $t$ (kW) |
| $P_{g,t}^{EL}$ / $P_{RG,t}^{EL}$ | Electricity purchased by electric load from grid/ RGs in period $t$ (kW) |
| $P_{g,t}^{HL}$ / $P_{RG,t}^{HL}$ | Electricity purchased by heat load from grid/ RGs in period $t$ (kW) |
| $P_{g,t}^{P2G}$ / $P_{RG,t}^{P2G}$ | Electricity purchased by P2G from grid/ RGs in period $t$ (kW) |
| $C_t$ | Capacities of energy storage devices in period $t$ (kWh) |
| $P_{ch,t}$ / $P_{dc,t}$ | Charging/ discharging power of energy storage devices in period $t$ (kW) |
| $Q_t^{P2G}$ | Natural gas produced by P2G in period $t$ (m³) |
| $Q_t^{MT}$ | Gas consumption volume of the MT in period $t$ (m³) |
| $P_t^{MT}$ / $H_t^{MT}$ | Output power/ heat of the MT in period $t$ (kW) |
| $Q_{g,t}^{MT}$ / $Q_{p2g,t}^{MT}$ | MT consumes natural gas volume from gas grid/ P2G in period $t$ (m³) |
| $Q_{g,t}^{GL}$ | Gas volume purchased from gas grid in period $t$ (m³) |
| $R_t^{grid}$ / $R_t^{ESD}$ | Spinning reserves of the power grid/ ESD in period $t$ (kW) |
| $P_t^{EV}$ | Charging power of EVs in period $t$ (kW) |
| $P_t^s$ | Power of renewable curtailment in period $t$ (kW) |
| $E(P_t^{RG})$ | The expectation of joint outputs of RGs in period $t$ (kW) |
| $E(P_t^{PV})$ / $E(P_t^{WT})$ | The expectation of output of PV/ WT in period $t$ (kW) |
| $\theta_t$ | P2G operation state |
| $T_n$ | Charging time of EV $n$ |
| $C_{n,t}^{EV}$ | Stored energy of the EV $n$ in period $t$ (kWh) |

**Parameters**

This work is supported by the Natural Science Foundation of Jilin Province, China under Grant No. 2020122349JC. (Corresponding author: Yang Li)

Y. Li, B. Wang, J. Li and G. Li are with the School of Electrical Engineering, Northeast Electric Power University, Jilin 132012, China (e-mail: liyang@neepu.edu.cn; 2201900118@neepu.edu.cn; tytrytytgr@gmail.com; lgq@neepu.edu.cn).

Z. Yang is with the State Grid Beijing Electric Power Company, Beijing 100032, China (e-mail: 1678084931@qq.com).

| Symbol | Description |
|---|---|
| $\alpha$ | Confidence level (%) |
| $q$ | Discrete step size (kW) |
| $P_{load,t}^0$ | Initial electric load in period $t$ (kW) |
| $Q_{load,t}^0$ | Initial gas load in period $t$ (m³) |
| $H_{load,t}^0$ | Initial heat load in period $t$ (kW) |
| $\alpha_{TSE}/\alpha_{IE}$ | Ratios of TSE/ IE to total electric load in period $t$ (%) |
| $\alpha_{TSQ}/\alpha_{IQ}$ | Ratios of TSQ/ IQ to total gas load in period $t$ (%) |
| $M$ | Human energy metabolism rate (W/m²) |
| $I_{cl}$ | The thermal resistance of clothing (m²·  /W) |
| $T_s$ | The average temperature of the human skin in a comfortable state ( ) |
| $T_{in,t}/T_{out,t}$ | Indoor/ outdoor temperature ( ) |
| $K$ | Comprehensive heat transfer factor (W/m²) |
| $F$ | Building surface area (m²) |
| $V$ | Building volume (m³) |
| $c_{air}$ | Heat capacity of indoor air (kJ·kg⁻¹·  ⁻¹) |
| $\rho_{air}$ | The density of indoor air (kg·m⁻³) |
| $\eta^{EB}/\eta^{P2G}$ | The conversion efficiency of EB/ P2G |
| $\eta_{ch}/\eta_{dc}$ | Charging/ discharging efficiencies of energy stored devices |
| $k_{loss}$ | Loss rate factor of energy stored devices |
| HHV | The calorific value of natural gas (kW/m³) |
| $\eta_e^{MT}/\eta_h^{MT}$ | Electricity power/ heat generation efficiency coefficients of MT |
| $\eta_{loss}^{MT}$ | Heat loss coefficient of MT |
| $\omega_{g,t}^P/\omega_{g,t}^Q$ | TOU power price/ natural gas price (¥) |
| $\delta_t^{grid}/\delta^{ESD}$ | Spinning reserve cost of the power grid/ ESD (¥/kWh) |
| $\varphi_l$ | Maintenance cost of device $l$ (¥/kW) |
| $\mu_{g,j}^P/\mu_{g,j}^Q$ | Emission coefficient of pollutant $j$ from the electricity/ natural gas purchased |
| $\mu^{P2G}$ | Coefficient of $CO_2$ absorption of the P2G |
| $\vartheta_j$ | Penalty price of the pollutant $j$ (¥) |
| $\gamma_{IE}/\gamma_{TSE}/\gamma_{CH}/\gamma_{IQ}/\gamma_{TSQ}$ | Unit compensation cost for IE/ TSE/ CH/ IQ/ TSQ (¥/kWh) |
| $P_{ch,max}/P_{dc,max}$ | Maximum charging/ discharging power of energy storage devices (kW) |
| $C_{max}/C_{min}$ | Maximum/ minimum capacity of energy storage devices (kWh) |
| $C_0/C_{T_{end}}$ | Starting/ ending capacity in one scheduling cycle (kWh) |
| $H_{max}^{EB}$ | Maximum heat power output of the EB (kW) |
| $P_{max}^{P2G}/P_{min}^{P2G}$ | Maximum/ minimum input power of P2G (kW) |
| $\Delta Q_{min}^{MT}/\Delta Q_{max}^{MT}$ | Lower/ upper limits of the climbing ability of MT (m³) |
| $P_{max}^{grid}$ | Maximum power provided by the grid for CIES (kW) |
| $S_{e,n}^{EV}$ | Expected SOC of EV $n$ (%) |
| $S_{real,t,n}^{EV}$ | Real state SOC at the start of charging of EV $n$ (%) |
| $C_{max,n}^{EV}/C_{min,n}^{EV}$ | Maximum/ minimum battery capacity of EV $n$ (kWh) |
| $W_{100}$ | Power consumption of 100 kilometers (kWh) |
| $P_{ch,n}^{EV}/\eta_{ch}^{EV}$ | Rated charging power/ efficiency of EV $n$ (kW) |
| $P_{ch,max}^{EV}$ | Upper bound of the charging power of all EVs (kW) |
| $N$ | Total number of EVs |

## I. Introduction

WITH environmental pollution and the exhaustion of traditional fossil energy have become increasingly pressing. The vigorous development of renewable generation (RG) and improvements in energy efficiency are commonly applied by the international community in an attempt to overcome such problems [1], [2]. The coupling degree between electrical and natural gas systems increases with the development of technologies and scales of power to gas (P2G) and micro-gas turbine (MT). Moreover, integrated energy systems (IESs) are able to achieve multi-energy complementation and collaboration via the coupling of independent energy systems, which consequently reduces operating costs and improves integrated energy efficiency [3], [4]. The modeling and optimization of a community integrated energy system (CIES), coupled with the energy carriers power, heat, and natural gas, provides a more economical and efficient option. Thus, an effective optimal scheduling strategy is in urgent demand.

At present, some studies have been performed focusing on the scheduling of IESs. Ref. [5] reveals a reduction in the wind power curtailment following the introduction of a heat storage device (HSD) and electric boiler (EB) to supply the heat-electric system. Ref. [6] proposes a scheduling strategy to integrate the random outputs of RG via the application of the charging flexibility of electric vehicles (EVs). Ref. [7] proposes a low-carbon scheduling model, in which the P2G device was employed to reduce $CO_2$ emission. However, the aforementioned studies failed to fully exploit the interaction ability of demand-side resources. Demand response in IESs has been considered a key measure to stimulate the interaction between demand-side resources and renewable energy [8]. The authors of [9] establish an integrated demand response (IDR) program considering the composition of electrical and heat loads. Ref. [10] models the interruptible, adjustable, and shiftable loads in a smart residential community. An IDR-based energy hub program is proposed in Ref. [11], whereby the electricity is switched to natural gas during peak hours. The increase in penetration of RGs (e.g., photovoltaic (PV) and wind turbine (WT)) results in further uncertainties, thereby posing new challenges for IES optimal scheduling.

Numerous approaches have been proposed to deal with the uncertainties associated with IES scheduling [12], including robust optimization [13-15], scenario-based method [16], [17] and chance-constrained programming (CCP) [18], [19]. The authors of [13] adopt robust optimization to deal with the uncertainty of electricity price considering IDR and EVs. In [14], the operating costs of multi-energy micro-grids are minimized under uncertainties via a two-stage robust optimization model. Ref. [15] developed a robust IES capacity model that considers both the demand response and thermal comfort. Robust optimization has proven to be highly promising in the analysis of the worst-case uncertainty scenario, however the solution is often conservative as it is a hedge against the worst-case realization [20].

A scenario-based method is another effective mathematical tool to handle uncertainties. In [16], a two-stage scenario-based stochastic programming is used for modeling the energy management problem of virtual power plants. Ref. [17] developed an optimization framework based on a hybrid scenario-based/interval/information gap decision theory method for energy hubs under uncertainties of load, energy prices, and renewable sources. However, the optimization results of such kind of scenario-based methods are heavily dependent on the quality of scenario generation and scenarios reduction method.

CCP has recently gained a great deal of attention in addressing uncertainties. Ref. [18] proposed a CCP-based hierarchical stochastic energy management strategy for interconnected microgrids with consideration of the uncertainties. In [19], a CCP-based optimization model is developed for the optimal operation of a hydro-PV system. Such CCP-based approaches can achieve a trade-off between system reliability and economy by setting a proper confidence level of chance constraints.

Table I compares our proposed approach with the methods in the current literature, highlighting the unique features of our method. Despite the pioneering research performed by the aforementioned studies, there are still the following open problems. (1) IDR enabled CIES is insufficiently considered. In particular, the demand response potential of natural gas and heat loads remain to be fully explored. (2) The majority of previous studies fail to consider the spinning reserve induced by the prediction errors between the RG predictions and the actual outputs. (3) How to deal with multiple renewable uncertainties efficiently is still a challenging problem in scheduling. (4) Research on the optimal scheduling of CIES that simultaneously considers IDR, spinning reserves, EVs, and multiple RG uncertainties is limited.

TABLE I
COMPARISON OF THE PROPOSED APPROACH WITH RELATED STUDIES

| Ref | Renewable Uncertainties | | IDR | | | EV | Spinning reserves |
|---|---|---|---|---|---|---|---|
| | WT | PV | Electricity | Heat | Gas | | |
| [13] | × | × | | | | × | × |
| [14] | × | | × | × | × | × | × |
| [15] | | × | | | | × | × |
| [16] | | | × | × | | × | × |
| [17] | | | | × | | × | × |
| [18] | | | × | × | × | | × |
| [19] | × | | × | × | × | × | × |
| Proposed | | | | | | | |

To address these issues, this study proposes an optimal scheduling approach for integrated demand response-enabled CIES with uncertain renewable generations. The key contributions of our work are summarized as follows:

(1) Aiming to seek the minimum operating costs, we propose a scheduling strategy based on chance-constrained programming (CCP) for integrated demand response-enabled CIES in uncertain environments. Here, an IDR program is employed to fully explore the potential of electricity-gas-heat flexible loads and electric vehicles. Due to the multiple renewable uncertainties, the spinning reserves provided by an electricity storage device (ESD) and a power grid are constructed in the form of chance constraints. Moreover, P2G and MT, as links of multi-energy carriers, are depicted to participate in the CIES scheduling.

(2) Sequence operation theory (SOT) is adopted to convert the chance constraints into their deterministic equivalent form. A solvable mixed-integer linear programming (MILP) model is then determined via the linearization methods, proving to be efficient in solving the CCP model.

(3) Numerical simulation results on the CIES in North China demonstrate the ability of the proposed method to reduce CIES operating costs by coordinating IDR and renewable uncertainties. In addition, the CIES operation can achieve a balance between economy and system reliability by setting a proper confidence level of spinning reserve constraints. The P2G and MT are able to enhance the system operational flexibility and the user comprehensive satisfaction.

## II. STRUCTURE MODELING OF CIES

To clearly demonstrate the CIES' physical model, Fig. 1 presents a schematic diagram of the CIES. The main components of the CIES include PV and WT units, an EB, EV charging station, P2G, an MT unit, energy storage devices, and loads. Note that hydrogen is an intermediate product in the P2G process in this study. Hydrogen can also be used as a source for hydrogen fuel cell vehicles in real applications.

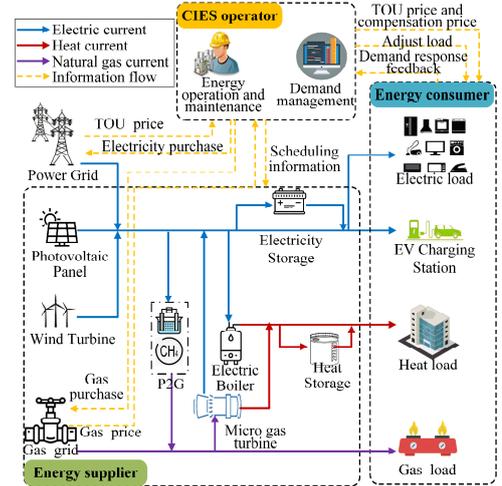

Fig. 1. Schematic diagram of the CIES.

### A. Probabilistic Wind Turbine Model

Studies have shown that WT output power $P^{WT}$ depends on wind speed and WT rated output [21]. Moreover, the wind speed obeys the Weibull distribution. Thus, the probability density function (PDF) of $P^{WT}$ can be formulated as:

$$f_w(P^{WT}) = \begin{cases} (khv_{in}/\varepsilon P_r)\left[((1+hP^{WT}/P_r)v_{in})/\varepsilon\right]^{k-1} \times \\ \quad \exp\left\{-\left[((1+hP^{WT}/P_r)v_{in})/\varepsilon\right]^k\right\}, \\ \quad P^{WT} \in [0, P_r] \\ 0, \quad otherwise \end{cases} \quad (1)$$

where $\varepsilon$ and $k$ are the scale factor and the shape factor; $p_r$ is the rated WT output; $h = (v_r/v_{in}) - 1$; $v_{in}$ and $v_r$ are the cut-in and rated wind speeds, respectively. More details of the probabilistic WT model can be found in [22].

### B. Probabilistic Photovoltaic Model

Existing researches suggest that the solar irradiance approximately obeys the Beta distribution, while the PV power output $P^{PV}$ has a linear relationship with solar irradiance [23], [24]. Thus, the PDF of $P^{PV}$ is

$$f_p(P^{PV}) = \frac{\Gamma(\lambda_1) + \Gamma(\lambda_2)}{\Gamma(\lambda_1)\Gamma(\lambda_2)} \left(\frac{P^{PV}}{P^{PV}_{max}}\right)^{\lambda_1-1} \left(1 - \frac{P^{PV}}{P^{PV}_{max}}\right)^{\lambda_2-1} \quad (2)$$

where $P^{PV}_{max}$ is the maximum value of $P^{PV}$; $\lambda_1$ and $\lambda_2$ are the shape factors; $\Gamma$ represents a Gramma function; $\xi$ and $\xi_{max}$ are the actual and maximum solar irradiance, respectively.

### C. Electric Vehicle Model

In order to simplify the analysis procedure, we assume that the charging commences when the EV final return ends. The PDFs of the final return time and daily mileage can be described as (3) and (4), respectively [25], [26]:

$$f_{EV}(t) = \begin{cases} \frac{1}{\sqrt{2\pi}\sigma_s} \exp[-\frac{(t+24-\mu_s)^2}{2\sigma_s^2}], & 0 < t \le \mu_s - 12 \\ \frac{1}{\sqrt{2\pi}\sigma_s} \exp[-\frac{(t-\mu_s)^2}{2\sigma_s^2}], & \mu_s - 12 < t \le 24 \end{cases} \quad (3)$$

$$f_d(x) = \frac{1}{\sqrt{2\pi}x\sigma_d} \exp\left[-\frac{(\ln x - \mu_d)^2}{2\sigma_d^2}\right] \quad (4)$$

where $\mu_s$ and $\sigma_s$ are the mean and standard deviation of the time for EV to arrive at the charging station, respectively; $\mu_d$ and $\sigma_d$ are the mean and standard deviation of the daily mileage.

Based on the daily mileage of EV $n$, the SOC at the start of charging can be calculated by

$$S^{EV}_{real,t,n} = (S^{EV}_{e,n} - \frac{W_{100}x_n}{100 C^{EV}_{max,n}}) \times 100\% \quad (5)$$

The charging time of EV $n$ is calculated as:

$$T_n = (S^{EV}_{e,n} - S^{EV}_{real,t,n}) \frac{C^{EV}_{max,n}}{P^{EV}_{ch,n}\eta^{EV}_{ch}} \quad (6)$$

The EV daily driving distance and starting charging time are independent. We employ the Monte Carlo method to derive the charging load of each EV by simulating the corresponding driving distance and starting charging time. This allows us to obtain the total charging load by superimposing the load of each EV. Fig. 2 depicts the EV daily load under disorderly charging.

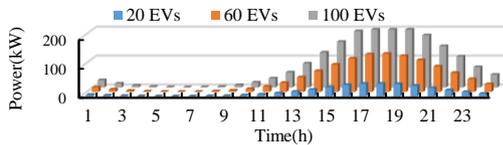

Fig.2 Daily load of EVs under disorderly charging

### D. Modeling of Load

#### 1) Electric Load Model

We divide the electric load into fixed and flexible loads based on the nature of the demand-side loads. Flexible loads include time-shiftable electric load (TSE) and interruptible electric load (IE) [23]. Since the load shift and interruption inevitably affect user experience, a certain subsidy will generally be provided to users [27]. The aggregated electric load participates in CIES scheduling and is described as:

$$P_{load,t} = P^0_{load,t} + P^{TSE}_t - P^{IE}_t \quad \forall t \quad (7)$$

$$-\alpha_{TSE} P^0_{load,t} \le P^{TSE}_t \le \alpha_{TSE} P^0_{load,t} \quad \forall t \quad (8)$$

$$\sum_{t=1}^{T} P^{TSE}_t = 0 \quad (9)$$

$$P^{IE}_t \le \alpha_{IE} P^0_{load,t} \quad \forall t \quad (10)$$

#### 2) Gas Load Model

Similar to the electric load model, the natural gas DR load can be modeled according to (11)-(14).

$$Q_{load,t} = Q^0_{load,t} + Q^{TSQ}_t - Q^{IQ}_t \quad \forall t \quad (11)$$

$$-\alpha_{TSQ} Q^0_{load,t} \le Q^{TSQ}_t \le \alpha_{TSQ} Q^0_{load,t} \quad \forall t \quad (12)$$

$$\sum_{t=1}^{T} Q^{TSQ}_t = 0 \quad (13)$$

$$Q^{IQ}_t \le \alpha_{IQ} Q^0_{load,t} \quad \forall t \quad (14)$$

#### 3) Heat Load Model

This work considers the building thermal inertia in heat load models. Since the user's perception of temperature comfort has a certain flexibility, the heat load can be reduced within an acceptable thermal comfort range for users. Thus, we can define the actual heat load $H_{load,t}$ as:

$$H_{load,t} = H^0_{load,t} - H^{CH}_t \quad (15)$$

To quantify the acceptable thermal comfort range of users, the Predictive Mean Vote (PMV) index is introduced [28]:

$$PMV = 2.43 - \frac{3.76(T_s - T_{in,t})}{M(I_{cl} + 0.1)} \quad (16)$$

According to ISO 7730 standard, we set the PMV limits as follows:

$$\begin{cases} |PMV| \le 0.9 & [1:00\text{-}7:00], [20:00\text{-}24:00] \\ |PMV| \le 0.5 & [8:00\text{-}19:00] \end{cases} \quad (17)$$

Moreover, based on our previous work [29], the heating power stored in a building is calculated by

$$H^0_{load,t} = \frac{[T_{in,t} - T_{out,t}] + \frac{K \cdot F}{c_{air} \cdot \rho_{air} \cdot V} \cdot \Delta t \cdot [T_{in,t} - T_{out,t}]}{\frac{1}{K \cdot F} + \frac{1}{c_{air} \cdot \rho_{air} \cdot V} \cdot \Delta t} \quad (18)$$

To comprehensively measure the impact of IDR on user experience, a user comprehensive satisfaction is designed as

$$m_{s,t} = \frac{\left(1 - \frac{|P_{load,t} - P^0_{load,t}|}{P^0_{load,t}}\right) + \left(1 - \frac{|H_{load,t} - H^0_{load,t}|}{H^0_{load,t}}\right) + \left(1 - \frac{|Q_{load,t} - Q^0_{load,t}|}{Q^0_{load,t}}\right)}{3} \times 100\% \quad (19)$$

### E. Electric Boiler Model

As a kind of electro-thermal coupling unit, EBs yield no pollutant emissions. The output model of EBs is

$$H^{EB}_t = \eta^{EB} P^{EB}_t = \eta^{EB}(P^{HL}_{g,t} + P^{HL}_{RG,t}) \quad \forall t \quad (20)$$

## F. Energy Storage Device Model

The energy storage devices described in this study can be grouped into two types, namely, ESD and HSD. The energy storage device model is formulated as:

$$C_{t+1} = (1-k_{loss})C_t + (\eta_{ch}P_{ch,t} - P_{dc,t}/\eta_{dc})\Delta t \quad \forall t \quad (21)$$

## G. Power to Gas Model

P2G technology includes two chemical processes. First, an electrolyzer is used to electrolyze $H_2O$ into $H_2$ and $O_2$, and then $H_2$ and $CO_2$ are synthesized into $CH_4$ through the Sabatier reaction.

$$\begin{cases} 2H_2O \rightarrow 2H_2 + O_2 \\ CO_2 + 4H_2 \rightarrow CH_4 + 2H_2O \end{cases} \quad (22)$$

The relationship between natural gas produced by P2G and electricity consumption is given by

$$Q_t^{P2G} = \frac{\eta^{P2G}P_t^{P2G}}{HHV} = \frac{\eta^{P2G}(P_{g,t}^{P2G}+P_{RG,t}^{P2G})}{HHV} \quad \forall t \quad (23)$$

## H. Micro-Gas Turbine Model

The relationship between input power and output power of the MT can be described as:

$$P_t^{MT} = \eta_e^{MT}Q_t^{MT}HHV \quad \forall t \quad (24)$$

$$H_t^{MT} = (1-\eta_e^{MT}-\eta_{loss}^{MT})Q_t^{MT}HHV \quad \forall t \quad (25)$$

$$Q_t^{MT} = Q_{g,t}^{MT} + Q_{p2g,t}^{MT} \quad \forall t \quad (26)$$

## III. OPTIMAL SCHEDULING MODEL OF CIES

The proposed scheduling approach of CIES, which aims to seek the minimum operating costs. In the following, we describe each of its components.

### A. Objective Function

The objective function pursues the minimization of the total operating cost, which is formulated as:

$$\min OF = C_1 + C_2 + C_3 + C_4 + C_5$$

$$C_1 = \sum_{t=1}^{T}\left[\omega_{g,t}^P\left(P_{g,t}^{EL}+P_{g,t}^{HL}+P_{g,t}^{P2G}\right)+\omega_{g,t}^Q Q_{g,t}^{GL}\right]\Delta t$$

$$C_2 = \sum_{t=1}^{T}\left(\delta_t^{grid}R_t^{grid}+\delta^{ESD}R_t^{ESD}\right)\Delta t$$

$$C_3 = \sum_{t=1}^{T}\sum_{l=1}^{L}\left(\varphi_l P_{l,t}\right)\Delta t \quad (27)$$

$$C_4 = \sum_{t=1}^{T}\sum_{j=1}^{J}\vartheta_j\begin{bmatrix}\mu_{g,j}^P\left(P_{g,t}^{EL}+P_{g,t}^{HL}+P_{g,t}^{P2G}+R_t^{grid}\right)\\+\mu_{g,j}^Q Q_{g,t}^{GL}HHV-\mu_j^{P2G}P_t^{P2G}\end{bmatrix}\Delta t$$

$$C_5 = \sum_{t=1}^{T}\begin{bmatrix}\gamma_{IE}P_t^{IE}-\left(\gamma_{TSE}\times\min\{P_t^{TSE},0\}\right)+\gamma_{CH}H_t^{CH}\\+\gamma_{IQ}Q_t^{IQ}-\left(\gamma_{TSQ}\times\min\{Q_t^{TSQ},0\}\right)\end{bmatrix}\Delta t$$

where $C_1$ is the energy transaction cost; $C_2$ is the spinning reserve cost; $C_3$ is the maintenance cost; $C_4$ is the environmental cost; $C_5$ is the IDR compensation cost.

### B. Constraint Conditions

#### 1) Energy Balance Constraints

To avoid the imbalance between the supply and demand caused by the excessive RGs outputs, CIES must equip load shedding $P_t^s$. Therefore, the energy balance constraints are

$$P_{g,t}^{EL}+P_{RG,t}^{EL}+P_{dc,t}^{ESD}+P_t^{MT} = P_{load,t}+\sum_{n=1}^{N}P_{ch,n,t}^{EV}+P_{ch,t}^{ESD}+P_t^s \quad \forall t \quad (28)$$

$$H_t^{EB}+H_{dc,t}^{HSD}-H_{ch,t}^{HSD}+H_t^{MT} = H_{load,t} \quad \forall t \quad (29)$$

$$Q_{g,t}^{GL}+Q_t^{P2G}-Q_t^{MT} = Q_{load,t} \quad \forall t \quad (30)$$

#### 2) Renewable Consumption Constraint

The expected value of the RGs joint outputs should meet the following power balance constraint:

$$E(P_t^{RG}) = P_{RG,t}^{EL}+P_{RG,t}^{P2G}+P_{RG,t}^{HL}+P_t^s \quad \forall t \quad (31)$$

#### 3) Energy Storage Device Constraints

The charging and discharging power are constrained as:

$$\begin{cases} 0 \leq P_{dc,t} \leq P_{dc,\max} \\ 0 \leq P_{ch,t} \leq P_{ch,\max} \end{cases} \quad \forall t \quad (32)$$

Furthermore, to ensure the same initial conditions in each scheduling cycle, the starting capacity should be equal to the ending capacity:

$$\begin{cases} C_{\min} \leq C_t \leq C_{\max} \\ C_0 = C_{\min} = C_{T_{end}} \end{cases} \quad \forall t \quad (33)$$

#### 4) Electric Boiler Constraint

The safe operation of the EB requires that its output power satisfies the following condition:

$$0 \leq H_t^{EB} \leq H_{\max}^{EB} \quad \forall t \quad (34)$$

#### 5) P2G Operating Constraints

The input power and ramp rate of the P2G are constrained by (35) and (36), respectively,

$$\begin{cases} P_{\min}^{P2G}\theta_t \leq P_t^{P2G} \leq P_{\max}^{P2G}\theta_t \\ \theta_t = 1 \ on; \ \theta_t = 0 \ off \end{cases} \quad \forall t \quad (35)$$

$$\Delta P_{\min}^{P2G}\theta_t \leq P_t^{P2G}-P_{t-1}^{P2G} \leq \Delta P_{\max}^{P2G}\theta_t \quad \forall t \quad (36)$$

#### 6) MT Operating Constraints

The input of MT obeys the following inequalities:

$$\begin{cases} Q_{\min}^{MT}\psi_t \leq Q_t^{MT} \leq Q_{\max}^{MT}\psi_t \\ \psi_t = 1 \ on; \ \psi_t = 0 \ off \end{cases} \quad \forall t \quad (37)$$

$$\Delta Q_{\min}^{MT}\psi_t \leq Q_t^{MT}-Q_{t-1}^{MT} \leq \Delta Q_{\max}^{MT}\psi_t \quad \forall t \quad (38)$$

#### 7) EVs Constraints

The EV charging station power should not exceed the maximum allowable power of the CIES,

$$0 \leq \sum_{n=1}^{N}P_{ch,n,t}^{EV} \leq P_{ch,\max}^{EV} \quad (39)$$

The EV capacity when charging must satisfy the following:

$$C_{n,t}^{EV} = C_{n,t-1}^{EV}+P_{ch,n}^{EV}\eta_{ch}^{EV}\Delta t \quad (40)$$

$$C_{\min,n}^{EV} \leq C_{n,t}^{EV} \leq C_{\max,n}^{EV} \quad (41)$$

#### 8) Spinning Reserve Constraints

This study considers the ability of the ESD and grid to participate in the provision of reserve services. Considering extreme situations where RGs outputs may be zero, the spinning reserve constraint is modeled in a chance constraint form [22]:

$$(P_{g,t}^{EL} + P_{g,t}^{HL}) + R_t^{grid} \leq P_{\max}^{grid} \quad \forall t \quad (42)$$

$$\begin{cases} R_t^{ESD} \leq \eta_{dc}(C_t^{ESD} - C_{\min}^{ESD})/\Delta t \\ R_t^{ESD} \leq P_{dc,\max}^{ESD} - P_{dc,t}^{ESD} \end{cases} \forall t \quad (43)$$

$$P_{rob}\{R_t^{grid} + R_t^{ESD} \geq E(P_t^{RG}) - P_t^{WT} - P_t^{PV}\} \geq \alpha \quad \forall t \quad (44)$$

## IV. MODEL CONVERSION AND SOLUTION

In this section, the chance constraint in the proposed scheduling model is converted into its deterministic equivalent form via the SOT. Then the equivalent model is transformed into a solvable MILP model by using linearization methods. Finally, solved by the CPLEX solver.

### A. Probabilistic Sequence of RG Outputs

The PV and WT outputs, with PDFs described in (1) and (2) respectively, are random variables in period $t$, and thus can be discretized to obtain probability sequences $a(i_{at})$ and $b(i_{bt})$. Taking PV output $P^{PV}$ as an example, the probabilistic sequence length $N_{at}$ of $P^{PV}$ is described as [29]:

$$N_{at} = [P_{\max,t}^{PV}/q] \quad (45)$$

where $[\cdot]$ is the ceiling function; $P_{\max,t}^{PV}$ is the maximum possible output value of the PV in period $t$; $q$ is the discrete step size; After discretization, there are a total of $N_{at}+1$ states, in which the output of the state $u_a$ is $u_a q (0 \leq u_a \leq N_{at})$, and the corresponding probability is $a(u_a)$. The PV output and its probabilistic sequence are shown in Table .

TABLE  
PV OUTPUT AND ITS CORRESPONDING PROBABILISTIC SEQUENCE

| Power/kW | 0 | $q$ | $2q$ | … | $u_a q$ | $N_a q$ |
|---|---|---|---|---|---|---|
| Probability | $a(0)$ | $a(q)$ | $a(2q)$ | … | $a(u_a)$ | $a(N_a)$ |

The probabilistic sequence $a(i_{at})$ of the PV output is calculated by using $f_p(P^{PV})$ as follows:

$$a(i_{at}) = \begin{cases} \int_0^{q/2} f_p(P^{PV}) dP^{PV}, & i_{at} = 0 \\ \int_{i_{at}q-q/2}^{i_{at}q+q/2} f_p(P^{PV}) dP^{PV}, & i_{at} > 0, i_{at} \neq N_{at} \\ \int_{i_{at}q-q/2}^{i_{at}q} f_p(P^{PV}) dP^{PV}, & i_{at} = N_{at} \end{cases} \quad (46)$$

The probabilistic sequence $b(i_{bt})$ of the WT output can be obtained via the same discretization method. Through the probability series of WT and PV in each period, $E(P_t^{RG})$ is determined as:

$$E(P_t^{RG}) = E(P_t^{PV}) + E(P_t^{WT}) = \sum_{u_{at}=0}^{N_{at}} u_{at} q a(u_{at}) + \sum_{u_{bt}=0}^{N_{bt}} u_{bt} q b(u_{bt}) \quad (47)$$

### B. Deterministic Conversion of Chance Constraints

There are two random variables with different distributions in (44). The prerequisite for the conversion from the chance constraint in (44) into its deterministic equivalence class is that the distribution of random variable $Z = P_t^{WT} + P_t^{PV}$ must be available. The distribution of variable $Z$ is as follows:

$$F_Z(z) = \int_{-\infty}^z \left[ \int_{-\infty}^\infty f_w(z-y) f_p(y) dy \right] dz \quad (48)$$

However, the determination of the inverse transform $F_Z^{-1}(z)$ is a critical challenge due to the complex form of the PDFs listed above. Besides, there may exist multiple solutions during the transformation process [30], [31]. Through utilizing the SOT to handle the probability distribution of the variable $Z$, the transformation from chance constraint to its deterministic class can be successfully achieved.

The probabilistic sequence of the joint outputs $c(i_{ct})$ is

$$c(i_{ct}) = \sum_{i_{at}+i_{bt}=i_{ct}} a(i_{at}) b(i_{bt}), i_{ct} = 0,1,\ldots,N_{at}+N_{bt} \quad (49)$$

The probability sequence $c(i_{ct})$ of $E(P_t^{RG})$ with the step size $q$ and the length $N_{ct}(N_{ct} = N_{at} + N_{bt})$ is shown in Table .

TABLE  
PROBABILISTIC SEQUENCE OF RG JOINT OUTPUTS

| Power/kW | 0 | $q$ | $2q$ | … | $(N_{ct}-1)q$ | $N_{ct}q$ |
|---|---|---|---|---|---|---|
| Probability | $c(0)$ | $c(q)$ | $c(2q)$ | … | $c(N_{ct}-1)$ | $c(N_{ct})$ |

In order to solve (44), we define a new class of Boolean variables $Z_{u_{ct}}$:

$$Z_{u_{ct}} = \begin{cases} 1, & R_t^{grid} + R_t^{ESD} \geq E(P_t^{RG}) - u_{ct}q \ (u_{ct} = 0,1,\ldots,N_{ct}) \\ 0, & \text{otherwise} \end{cases} \quad (50)$$

Based on Table , (44) can be simplified as:

$$\sum_{m_c=0}^{N_{at}+N_{bt}} Z_{u_{ct}} c(u_{ct}) \geq \alpha, \quad \forall t \quad (51)$$

Thus, we can convert (44) into a deterministic equivalent form.

### C. Linearization Methods

#### 1) Piecewise Function Linearization

$Z_{u_{ct}}$ cannot be directly solved by using MILP, hence we transform (50) as follows:

$$\frac{[R_t^{grid} + R_t^{ESD} + u_{ct}q - E(P_t^{RG})]}{\chi} \leq Z_{u_{ct}} \leq 1 + \frac{[R_t^{grid} + R_t^{ESD} + u_{ct}q - E(P_t^{RG})]}{\chi} \quad (52)$$

$\forall t, u_{c,t} = 0,1,\ldots,N_{at}+N_{bt}$

where $\chi$ is a large positive number.

When $R_t^{grid} + R_t^{ESD} \geq E(P_t^{RG}) - u_{ct}q$, (52) is equivalent to $\tau \leq Z_{u_{ct}} \leq \tau + 1$ ($\tau$ is a very small positive number). Since $Z_{u_{ct}}$ is a 0-1 variable, it can only be equal to 1; otherwise, it is equal to 0. This implies that (52) has exactly the same meaning as (50).

## 2) Elimination of Minimum Operators

Auxiliary variables are introduced to eliminate the minimum operators in the term $C_5$ of (27). Taking $\min\{P_t^{TSE},0\}$ as an example, we assume that $g(P_t^{TSE}) = \min\{P_t^{TSE},0\}$. According to (8), $g(P_t^{TSE})$ can be described as:

$$g(P_t^{TSE}) = \begin{cases} P_t^{TSE}, & -\alpha_{TSE} P_{load,t}^0 \leq P_t^{TSE} \leq 0 \\ 0, & 0 < P_t^{TSE} \leq \alpha_{TSE} P_{load,t}^0 \end{cases} \quad (53)$$

We then employ continuous auxiliary variables $w_1$, $w_2$, $w_3$ and 0-1 auxiliary variables $z_1$, $z_2$, $z_3$ to convert (53) into the following linear form:

$$\begin{cases} g(P_t^{TSE}) = w_{1,t} g(-\alpha_{TSE} P_{load,t}^0) + w_{2,t} g(0) + w_{3,t} g(\alpha_{TSE} P_{load,t}^0) \\ \quad = -w_{1,t} \alpha_{TSE} P_{load,t}^0 \\ P_t^{TSE} = w_{1,t}(-\alpha_{TSE} P_{load,t}^0) + w_{3,t}(\alpha_{TSE} P_{load,t}^0) \\ z_{1,t} + z_{2,t} + z_{3,t} = 1 \quad z_{1,t}, z_{2,t}, z_{3,t} \in \{0,1\} \\ w_{1,t} + w_{2,t} + w_{3,t} = 1 \\ w_{1,t} \leq z_{1,t};\ w_{2,t} \leq z_{1,t} + z_{2,t};\ w_{3,t} \leq z_{2,t} + z_{3,t} \end{cases} \forall t \ (54)$$

The same method is employed to convert $\min\{Q_t^{TSQ},0\}$ into its linear form and thus, the deterministic equivalent form is reformulated as a solvable MILP equation.

### D. Solution Process

The specific solution processes are as follows:
Step 1: Establish the model of each device in the CIES.
Step 2: IDR mechanism considering electricity-gas-heat flexible loads and electric vehicle is designed.
Step 3: A CCP-based CIES scheduling model is established.
Step 4: The PDFs of the RG outputs and discrete step $q$ are inputted.
Step 5: The PDFs of the RG outputs are discretized and their probabilistic sequences are generated.
Step 6: The SOT is adopted to obtain the expected value of the RG joint outputs and their corresponding probability sequence.
Step 7: The chance constraint is converted into the deterministic form.
Step 8: The deterministic equivalent model is converted into the MILP form via the linearization methods.
Step 9: The system parameters are input into the framework.
Step 10: The reserve capacity confidence level is set.
Step 11: The CPLEX solver is employed to solve the model.
Step 12: If the stop criterion is satisfied, the process is terminated; otherwise, the confidence level is modified and step 11 is repeated.
Step 13: The scheduling strategy is obtained.

## V. CASE STUDY

In order to verify the effectiveness of our approach, we implement a CIES in North China as the testing system. All simulations are performed on a PC platform equipped with 2 Intel Core dual-core CPUs (2.8 GHz) and 16 GB RAM.

### A. Description of Testing System

As shown in Fig. 1, the key components of the CIES include renewable power generations, an EB, a P2G device, an MT unit, storage devices, and an EV charging station. Table IV provides the main parameters of the system [23], [28-33]. Other parameters are described in the following. WT parameters: $\varepsilon = 1.8$, $k = 10$, $v_{in} = 3$m/s, $v_r = 15$m/s, $P_r = 600$kW; PV parameters $\lambda_1 = 3$, $\lambda_2 = 5$, $P_{max}^{PV} = 360$kW; and essential EVs parameters $N = 60$, $W_{100} = 15$ kWh/100km, $C_{max}^{EV} = 30$kWh, $P_{ch,n}^{EV} = 15$kW, $P_{ch,max}^{EV} = 225$kW, $\mu_s = 17.6$, $\sigma_s = 3.4$, $\mu_d = 3.2$, $\sigma_d = 0.88$.

TABLE
MAIN PARAMETERS OF THE CIES

| Parameter | Value | Parameter | Value |
|---|---|---|---|
| $H_{max}^{EB}$(kW) | 300 | $C_{min}^{HSD}$(kWh) | 0 |
| $\eta^{EB}$ | 0.99 | $C_{max}^{HSD}$(kWh) | 160 |
| $P_{min}^{P2G}$(kW) | 100 | $P_{ch,max}^{HSD}, P_{dc,max}^{HSD}$(kW) | 60 |
| $P_{max}^{P2G}$(kW) | 500 | $\eta_{ch}^{HSD}, \eta_{ch}^{HSD}$ | 1 |
| $\Delta P_{min}^{P2G}$(kW) | -200 | $k_{loss}^{HSD}$ | 0.01 |
| $\Delta P_{max}^{P2G}$(kW) | 200 | $Q_{min}^{MT}$(m$^3$) | 10 |
| $\eta^{P2G}$ | 0.6 | $Q_{max}^{MT}$(m$^3$) | 40 |
| $P_{max}^{grid}$(kW) | 1000 | $\Delta Q_{min}^{MT}$(m$^3$) | -10 |
| $Q_{max}^{grid}$(m$^3$) | 80 | $\Delta Q_{max}^{MT}$(m$^3$) | 10 |
| $C_{min}^{ESD}$(kWh) | 40 | $\eta_e^{MT}$ | 0.4 |
| $C_{max}^{ESD}$(kWh) | 200 | $\eta_h^{MT}$ | 0.5 |
| $P_{ch,max}^{ESD}, P_{dc,max}^{ESD}$ (kW) | 60 | $\eta_{loss}^{MT}$ | 0.1 |
| $\eta_{ch}^{ESD}, \eta_{ch}^{ESD}$ | 0.9 | $K$(W·m$^{-2}$) | 0.5 |
| $k_{loss}^{ESD}$ | 0.001 | $F$(m$^2$) | 2400 |
| $\delta^{ESD}$(¥/kWh) | 0.14 | $V$(m$^3$) | 36000 |
| $\varphi_{PV}$(¥/kWh) | 0.025 | $c_{air}$(kJ·kg$^{-1}$·°C$^{-1}$) | 1.007 |
| $\varphi_{WT}$(¥/kWh) | 0.025 | $\rho_{air}$(kg·m$^{-3}$) | 1.2 |
| $\varphi_{EB}$(¥/kWh) | 0.032 | $\varphi_{MT}$(¥/kWh) | 0.012 |
| $\varphi_{ESD}$(¥/kWh) | 0.002 | $\delta_{IE}$(¥/kWh) | 0.5 |
| $\varphi_{HSD}$(¥/kWh) | 0.005 | $\delta_{TSE}$(¥/kWh) | 0.3 |
| $\varphi_{P2G}$(¥/kWh) | 0.007 | $\delta_{CH}$(¥/kWh) | 0.4 |
| $\delta_{TSQ}$(¥/m$^3$) | 0.7 | $\delta_{IQ}$(¥/m$^3$) | 3.5 |

Fig. 3 presents the daily temperature curve and energy price, while Fig. 4 depicts the expected outputs of WT and PV and the predicted load values. The parameters related to pollution emissions can be found in [3] and [32]. In order to simplify analysis, the ratios of flexible loads $\alpha_{TSE}$, $\alpha_{IE}$, $\alpha_{TSQ}$ and $\alpha_{IQ}$ are considered to be fixed at 10% [33].

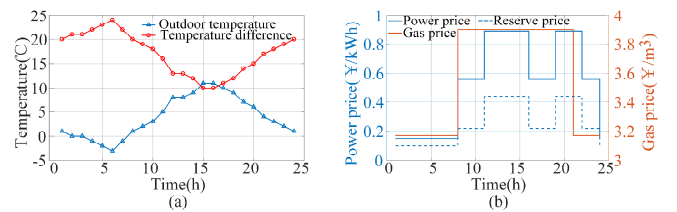

Fig. 3. Daily temperature curve and energy price.

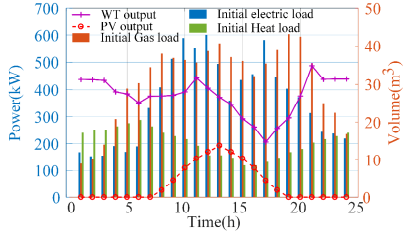

Fig. 4. WT and PV expected outputs and daily load variations.

*B. Results and Analysis*

We analyze the scheduling results of three operation scenarios to evaluate the performance of the proposed method. The three scenarios are defined as:

Scenario 1: Without IDR, P2G and MT participation.
Scenario 2: With IDR and without P2G and MT participation.
Scenario 3: **With IDR, P2G and MT participation (used in the proposed method).**

*1) Analysis of Confidence Levels*

To select a proper confidence level of spinning reserve, we analyze the reserve capacities in the CIES under five confidence levels in scenario 3. The results are shown in Fig. 5.

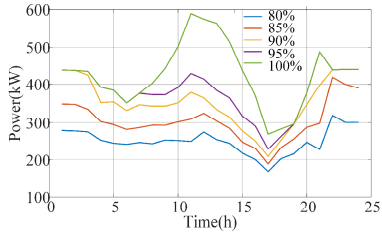

Fig. 5. Reserve capacities under different confidence levels.

Fig. 5 shows that as the confidence level increases, the required reserve capacity gradually increases, which improves the reliability of system operation. However, this will increase the operating costs of CIES. As a result, it is critical to choose a proper confidence level for balancing the reliability and economy of the system operation. According to our previous work [29], the confidence level is selected as $\alpha = 90\%$.

*2) Impact of IDR on Scheduling*

The energy supplies of each energy subsystem in scenarios 1 and 2 are compared, and the results are shown in Fig. 6.

Fig. 6 illustrates that there are significant differences in the energy dispatching scheme in scenarios 1 and 2. Specifically, regarding the electrical supply subsystems, a large amount of electricity will be purchased with a higher price during 9:00-10:00 and 15:00-20:00 while charging behaviors of EVs are at random in scenario 1, as shown in Fig. 6(a). Fig. 6(d) shows that the TSE plays a time-shifting role to shift 188.76kWh loads; the IE achieves the peak cutting effect and curtails 621.87kWh loads, and the EV charging plan is concentrated in the valley period. For the gas supply subsystem, the demand response of the flexible gas load implements the peak shaving and valley filling effects, and reduces the cost of purchasing gas in scenario 2, as shown in Fig. 6(e). For the heating subsystem, due to the influence of TOU energy prices and compensation mechanism, the heating power will be reduced during 8:00-23:00 in scenario 2, as shown in Fig. 6(f). The above analysis suggests the IDR can improve energy utilization efficiency and achieve the peak shaving and valley filling effects.

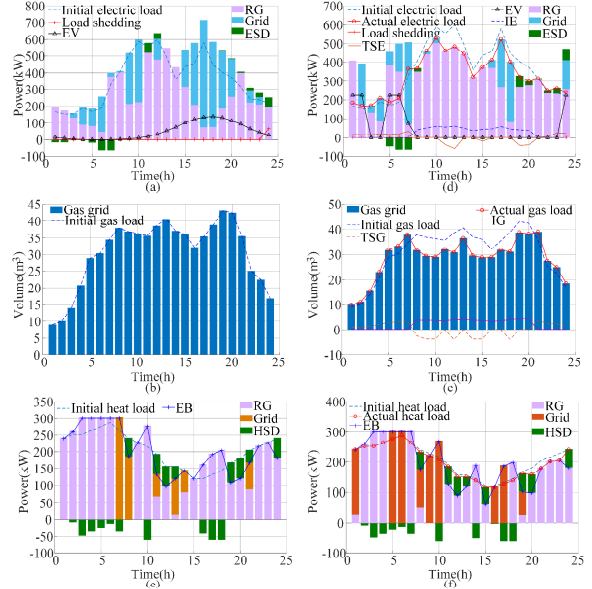

Fig. 6. Energy dispatching scheme in scenarios 1 [(a)-(c)] and 2 [(d)-(f)].

*3) Impact of P2G and MT on Scheduling*

In order to analyze the impact of P2G and MT on the schedule, we compare the scheduling schemes in scenarios 2 and 3. Fig. 7 depicts the scheduling results under scenario 3.

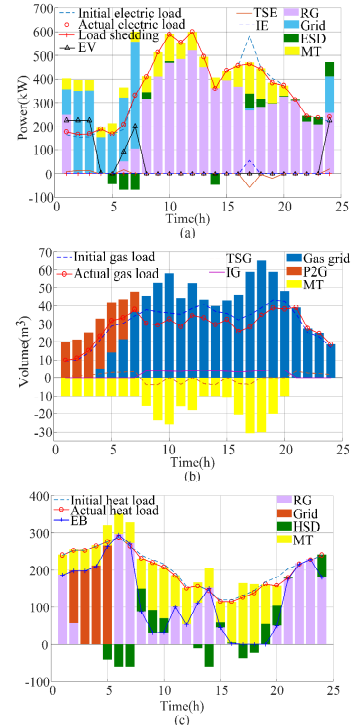

Fig.7. Energy dispatching scheme in scenario 3 for the (a) electrical, (b) gas supply, and (c) heating subsystems.

Fig. 7(a) suggests that the power purchased for the electrical supply subsystem in scenario 3 is concentrated during 0:00-7:00. It can be illustrated from Fig. 7(b) that P2G absorbs electric energy for energy conversion during 1:00-7:00, while MT participates in the system operation during 1:00-20:00. This shows that P2G and MT can improve the system operation flexibility by coupling different subsystems.

For the heating subsystem, by comparing Fig. 6(f) and Fig. 7(c), one can see that the power of the EB in scenario 3 is obviously lower than that in scenario 2 in most periods, and the electricity purchasing is mainly concentrated in period 2:00-5:00. This fact suggests that the MT actively participates in the subsystem operation due to the integrated demand response program, which provides additional flexibility for the system.

*4) Economic Analysis*

We perform a comparative analysis of the three scenarios to evaluate the economic costs and renewable consumption of the proposed method. The test results are listed in Table .

TABLE
ECONOMIC ANALYSIS RESULTS UNDER DIFFERENT SCENARIOS

| Item | Scenario 1 | Scenario 2 | Scenario 3 |
|---|---|---|---|
| Energy purchase cost (¥) | 5320.56 | 3746.97 | 3765.86 |
| Spinning reserve cost (¥) | 1937.95 | 1933.69 | 1918.84 |
| Maintenance cost (¥) | 416.91 | 413.49 | 419.69 |
| Environmental cost (¥) | 305.68 | 284.64 | 305.93 |
| IDR compensation cost (¥) | 0 | 620.43 | 287.36 |
| **Total operating costs (¥)** | **7981.1** | **6999.22** | **6697.68** |
| Renewable curtailment(kWh) | 66.1 | 0 | 0 |

Table shows that the total operating costs and renewable consumption vary with different scenarios. In scenario 1, the gas supply subsystem operates independently and the cost of purchasing energy for the CIES is the highest among all scenarios. In scenario 2, the IDR enables the CIES to reduce the total operating costs by 981.88¥ compared with scenario 1. In scenario 3, the IDR compensation and the total operating costs are reduced by 333.07¥ and 301.54¥ respectively compared to scenario 2. This indicates that compared with scenario 2, the CIES in scenario 3 provides a greater amount of energy at lower costs. This is attributed to the close coupling of all subsystems due to the participation of P2G and MT.

Most of the energy demand in the CIES is provided by external energy grids, and the energy purchase plan for each scenario is illustrated in Fig. 8.

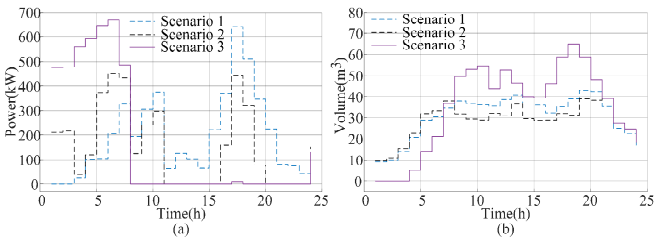

Fig.8. Purchased energy from energy grids under different scenarios.

Fig. 8 shows that energy purchase plans have obvious differences in different scenarios. To be specific, CIES purchases electricity and gas during valley periods and off-valley periods in scenario 3. The reason is that the MT is given as a priority to supply the required electric and thermal energies in off-valley periods; while in valley periods, the participation of P2G further improves the operational economy of the CIES.

*5) Satisfaction Analysis*

We analyze the impact of different schemes corresponding to scenarios 2 and 3 on user comprehensive satisfaction, the comparison results are shown in Fig. 9.

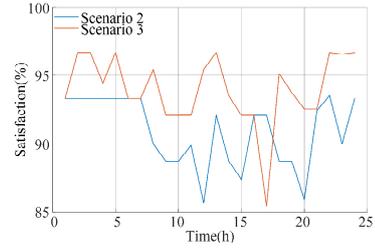

Fig.9. Comprehensive satisfaction index of scenarios 2 and 3.

Fig. 9 illustrates that comprehensive user satisfaction in scenario 3 generally exceeds that of scenario 2. This indicates that the participation of P2G and MT improves the CIES operational flexibility, enabling the system to better satisfy the load demands of users.

*C. Comparison with Hybrid Intelligent Algorithm*

In order to validate the performance of our solution method, we compare it with the Hybrid Intelligent Algorithm (HIA), which combines the particle swarm optimization (PSO) algorithm with Monte Carlo simulations. For the HIA, the population size is set to 100, and the rest parameters are taken from [22]. The comparison results are shown in Table .

TABLE
COMPARISON OF PROPOSED METHOD WITH HIA

| Confidence levels (%) | Proposed method | | HIA | |
|---|---|---|---|---|
| | Operating costs (¥) | Calculation time (s) | Operating costs (¥) | Calculation time (s) |
| 90 | 6697.68 | 2.31 | 6718.15 | 631.59 |
| 95 | 6918.21 | 2.32 | 7101.18 | 664.26 |
| 100 | 7317.66 | 2.41 | 7343.98 | 676.82 |

It can be seen from Table that the proposed method is superior to the HIA in terms of operating costs and computation time. 1) The operating costs calculated by using the presented method are lower than those of the HIA under different confidence levels; 2) The calculation time of the proposal is basically unchanged under different confidence levels and is remarkably less than that of the HIA.

## VI. CONCLUSION

In this study, an optimal scheduling strategy of integrated demand response-enabled CIES in uncertain environments is proposed. Based on the simulation results on a real-world CIES, the following conclusions can be summarized:

1) The electricity-gas-heat IDR mechanism is able to implement peak shaving and valley filling, improving the economy of system operation. By leveraging the IDR, the total system operating cost is decreased by 12.3% under a scheduling cycle with no renewable power curtailments.

2) As multi-energy carrier links, P2G and MT can enhance the system operational flexibility and user comprehensive satisfaction by taking advantage of multi-energy complementary benefits.

3) The spinning reserves provided by ESD and power grid are built in a form of chance constraints considering the multiple RGs uncertainties. By setting a proper confidence level, the CIES operation can achieve a balance between economy and reliability.

4) The developed solution approach surpasses the commonly used HIA due to its lower operating costs and higher calculation efficiency. By using the SOT and linearization methods, the original CCP-based scheduling framework is converted into a solvable MILP model and then solved by the CPLEX solver.

Note that this work ignores battery degradation, EVs interaction, and load uncertainty, while a more realistic scheduling scenario should consider these factors [34], [35]. Another interesting topic is incorporating integrated demand response and Game theory into the scheduling of IESs.